\documentclass[%
  reprint,
  showpacs,
  amsfonts,amssymb,amsmath,
  aps,
  pra,
]{revtex4-1}

\usepackage[ascii]{inputenc}
\usepackage[T1]{fontenc}
\usepackage[english]{babel}
\usepackage{graphicx}
\usepackage{microtype}
\usepackage[pdftex,colorlinks]{hyperref}
\hypersetup{%
  linkcolor=blue,%
  citecolor=blue,%
  urlcolor=blue,%
  pdftitle={Purity oscillations in Bose-Einstein condensates with balanced gain
            and loss},%
  pdfauthor={Dennis Dast, Daniel Haag, Holger Cartarius, G\"unter Wunner}%
}

\DeclareMathOperator{\trace}{tr}
\DeclareMathOperator{\asin}{asin}
\newcommand{\PT}{\mathcal{PT}}
\newcommand{\mrm}{\mathrm}
\newcommand{\ha}{\hat{a}}

\newcommand{\loss}{\mathrm{loss}}
\newcommand{\gain}{\mathrm{gain}}
\newcommand{\ket}[1]{|{#1}\rangle}
\newcommand{\mean}[1]{\langle{#1}\rangle}

\graphicspath{{./img/}}

\begin{document}

\title{Purity oscillations in Bose-Einstein condensates with balanced gain
       and loss}

\author{Dennis Dast}
\email[]{dennis.dast@itp1.uni-stuttgart.de}

\author{Daniel Haag}

\author{Holger Cartarius}

\author{G\"unter Wunner}

\affiliation{Institut f\"ur Theoretische Physik 1,
             Universit\"at Stuttgart, 70550 Stuttgart, Germany}

\date{\today}

\begin{abstract}
  In this work we present a new generic feature of $\PT$-symmetric
  Bose-Einstein condensates by studying the many-particle description of a
  two-mode condensate with balanced gain and loss.
  This is achieved using a master equation in Lindblad form whose mean-field
  limit is a $\PT$-symmetric Gross-Pitaevskii equation.
  It is shown that the purity of the condensate periodically drops to small
  values but then is nearly completely restored.
  This has a direct impact on the average contrast in interference experiments
  which cannot be covered by the mean-field approximation, in which a
  completely pure condensate is assumed.
\end{abstract}


\maketitle

Condensates with large numbers of atoms are accurately described in the
mean-field approximation by the Gross-Pitaevskii equation, however, if quantum
correlations are important the mean-field approach is no longer appropriate.
In particular, in the mean-field limit it is assumed that the condensate is
completely pure although in a real condensate the interaction of the particles
and the coupling to the environment will in general reduce the purity and
coherence~\cite{Ruostekoski98a}.
The coherence of the matter wave field, however, is the requirement for the
observation of a well-defined interference pattern.

A rather unusual behavior is the growth of the coherence in a system.
Yet, a collapse and revival of the matter wave field due to the interaction
between the particles has already been observed in the dynamical evolution of
the interference pattern by preparing a condensate in an optical lattice and
then ramping up the potential barrier to inhibit tunneling~\cite{Greiner02a}.
However, these revivals are damped by particle losses into the
environment~\cite{Pawlowski10a, Sinatra98a}.
On the other hand, it was shown in a two-mode system that taking dissipation
and phase noise into account can, if carefully prepared, yield a revival of the
purity before it eventually decays~\cite{Witthaut08a, Witthaut09a,
Witthaut11a}.

In this work we show that oscillations of the purity are a characteristic
feature of Bose-Einstein condensates subject to balanced gain and loss of
particles.
We demonstrate that the purity of this system does not simply decay.
Instead it drops periodically to small values and then is nearly completely
restored.
This has a direct impact on the average contrast measured in interference
experiments.

To do so we study a Bose-Einstein condensate on two lattice sites with an
influx of particles at one site and an outflux from the other site.
This serves as a model for a spatially extended double-well potential where
particles are removed or injected in the two wells.
The particle gain and loss is introduced via a master equation in Lindblad
form~\cite{Breuer02a} where the coherent dynamics is governed by the
Bose-Hubbard Hamiltonian~\cite{Jaksch98a} and the rate of the Lindblad
superoperators are balanced in such a way that the in- and outflux, at least
for small times, cancel each other out if half of the particles are at each
site~\cite{Dast14a}.

The mean-field limit of this master equation yields the Gross-Pitaevskii
equation where balanced particle gain and loss occurs as a $\PT$-symmetric
imaginary potential~\cite{Trimborn08a, Witthaut11a, Dast14a}.
Non-Hermitian but $\PT$-symmetric systems, i.e.\ systems whose Hamiltonians
commute with the combined action of the parity reflection and time reversal
operator, are known to support stationary solutions~\cite{Bender98a} under
certain conditions~\cite{Fernandez98a} and $\PT$-symmetric Bose-Einstein
condensates have been the subject of various studies~\cite{Graefe08b,
Cartarius12b, Dast13a, Fortanier14a}.
In these works stable stationary solutions, a rich dynamics, and a variety of
bifurcation scenarios were found.
Proposals for the experimental realization of a $\PT$-symmetric double well
exist by embedding it in a Hermitian four-well system~\cite{Kreibich13a,
Kreibich14a}.
Since the exchange of particles with the environment plays a crucial role in
these systems it cannot be expected that a mean-field approach is appropriate,
thus motivating the formulation of a many-particle description.

The master equation describing a Bose-Einstein condensate on two lattice sites
with balanced gain and loss was introduced in~\cite{Dast14a} and reads
\begin{subequations}
  \begin{gather}
    \frac{d}{dt}\hat\rho = -i [\hat H,\hat\rho]
    + \mathcal{L}_\loss\hat\rho + \mathcal{L}_\gain\hat\rho,
    \label{eq:mastereq}\\
    \hat H = -J(\ha_1^\dagger \ha_2 + \ha_2^\dagger \ha_1)
    + \frac{U}{2} (\ha_1^\dagger \ha_1^\dagger \ha_1 \ha_1
    + \ha_2^\dagger \ha_2^\dagger \ha_2 \ha_2),
    \label{eq:bhhamiltonian}\\
    \mathcal{L}_\loss \hat\rho = -\frac{\gamma_\loss}{2}
    (\ha_1^\dagger \ha_1 \hat\rho + \hat\rho \ha_1^\dagger \ha_1
    - 2 \ha_1 \hat\rho \ha_1^\dagger),
    \label{eq:liouvillianloss}\\
    \mathcal{L}_\gain \hat\rho = -\frac{\gamma_\gain}{2}
    (\ha_2 \ha_2^\dagger \hat\rho + \hat\rho \ha_2 \ha_2^\dagger
    - 2 \ha_2^\dagger \hat\rho \ha_2),
    \label{eq:liouvilliangain}
  \end{gather}
  \label{eq:completemaster}%
\end{subequations}
where $\ha_j^\dagger$ and $\ha_j$ are the bosonic creation and
annihilation operators, respectively.
Master equations are routinely used to describe phase noise and both feeding
and depleting of a Bose-Einstein condensate~\cite{Anglin97a, Ruostekoski98a}.
In~\cite{Dast14a} it was shown that the balanced gain and loss in the master
equation correctly reproduces characteristic properties of its $\PT$-symmetric
mean-field limit such as the in-phase pulsing between the lattice sites, and
thus can describe the underlying process of effective non-Hermitian
$\PT$-symmetric potentials.
Comparing the time evolution of expectation values such as the particle number
showed that there is an excellent agreement between the results of the master
equation with balanced gain and loss and the $\PT$-symmetric Gross-Pitaevskii
equation.
Here we show that there is in fact a crucial difference by analyzing the purity
of the condensate and the average contrast measured in interference
experiments.

The Bose-Hubbard Hamiltonian~\cite{Jaksch98a} in Eq.~\eqref{eq:bhhamiltonian}
describes the coherent dynamics of bosonic atoms in the lowest-energy Bloch
band of an optical lattice.
The tunneling rate between the two lattice sites is given by the parameter $J$,
and the strength of the on-site interaction by the parameter $U$.
We introduce the macroscopic interaction strength
\begin{equation}
  g=(N_0-1)U,
  \label{eq:macroscopicinteraction}
\end{equation}
which is used in the mean-field limit and depends on the initial amount of
particles $N_0$ in the system.

The underlying process of the particle loss localized at site 1 could be a
focused electron beam~\cite{Gericke08a, Wurtz09a} and the particle gain at site
2 may be induced by feeding from a second condensate~\cite{Robins08a} in a
Raman superradiance-like pumping process~\cite{Doring09a, Schneble04a,
Yoshikawa04a}.
The strength of the particle loss and gain is given by the parameters
$\gamma_\loss$ and $\gamma_\gain$, respectively.
The ratio
\begin{equation}
  \frac{\gamma_\gain}{\gamma_\loss} = 
    \frac{N_0}{N_0+2}
  \label{eq:gainlossratio}
\end{equation}
is chosen in such a way that for small times the gain and loss cancel each
other out if half of the particles are at each lattice site~\cite{Dast14a}.
In the following we use the notation $\gamma = \gamma_\loss$, while
$\gamma_\gain$ is always chosen such that it fulfills
relation~\eqref{eq:gainlossratio}.

The mean-field approximation is obtained in the limit $N_0 \to \infty$.
In this limit covariances are neglected and the condensate is pure.
For the two-mode system described by Eqs.~\eqref{eq:completemaster} this
yields the $\PT$-symmetric Gross-Pitaevskii equation~\cite{Dast14a}
\begin{subequations}
  \begin{align}
    i \frac{d}{dt} c_1 = - J c_2 + g |c_1|^2 c_1
      - i \frac{\gamma}{2} c_1,\\
    i \frac{d}{dt} c_2 = - J c_1 + g |c_2|^2 c_2
      + i \frac{\gamma}{2} c_2.
  \end{align}
  \label{eq:discreteGPE}%
\end{subequations}
The Gross-Pitaevskii equation has two $\PT$-symmetric stationary solutions
which we will refer to as the ground and excited state of the
system~\cite{Graefe12a, Dast14a}.
The parity reflection operator $\mathcal{P}$ is defined by its action on a
state vector $\mathcal{P}(c_1,\ c_2)^T=(c_2,\ c_1)^T$ and the time reversal
operator $\mathcal{T}$ is, as usual, a complex conjugation.
Their components read, up to a global phase,
\begin{equation}
  c_1 = \pm \frac{1}{\sqrt{2}}\exp{\left(\pm i \asin{
  \left(\frac{\gamma}{2J}\right)}\right)}, \quad
  c_2 = \frac{1}{\sqrt{2}}.
  \label{eq:stationaryStates}
\end{equation}
For $\gamma=0$ the ground state (positive signs in
Eq.~\eqref{eq:stationaryStates}) is symmetric and the excited state (negative
signs) antisymmetric.
These solutions only exist for $|\gamma| \leq 2J$.
Two additional decaying or growing $\PT$-broken solutions exist in the regime
$|\gamma| \geq \sqrt{4J^2 - g^2}$.
To avoid the influence of an instability introduced by the $\PT$-broken states
all calculations are done in the parameter regime of unbroken $\PT$ symmetry.

In this work we solve the master equation \eqref{eq:completemaster} via the
quantum jump method~\cite{Molmer93a, Johansson13a}.
We average over a certain amount of quantum trajectories till the results
converge.
In all calculations the initial state is a pure product state, which in the
mean-field limit is defined by two complex numbers $c = (c_1,\ c_2)^T$.
The many-particle state with $N_0$ particles that corresponds to this
mean-field state reads
\begin{equation}
  \ket{\psi} = \sum_{m=0}^{N_0}
  \sqrt{ \begin{pmatrix} N_0 \\ m \end{pmatrix} }
  c_1^{N_0-m}c_2^m \ket{N_0-m,m},
  \label{eq:productState}
\end{equation}
where $\ket{n_1, n_2}$ describes a Fock state with $n_j$ particles at lattice
site $j$~\cite{Dast14a}.
Furthermore in the following the tunneling rate is set to $J=1$.

The purity of the reduced single-particle density matrix $\sigma_{\mrm{red},jk}
= \mean{\ha_j^\dagger \ha_k}/\sum_i \mean{\ha_i^\dagger\ha_i}$ measures how
close the condensate is to a pure Bose-Einstein condensate~\cite{Penrose56a}
and is defined as
\begin{equation}
  P = 2 \trace \sigma_\mrm{red}^2 - 1 \ \in [0,1].
  \label{eq:purity}
\end{equation}
Only in a pure condensate, i.e.\ $P=1$, all atoms are in the same
single-particle state as assumed for the Gross-Pitaevskii equation and the
eigenvalues of the single-particle density matrix are one and zero.
For smaller values of $P$ there is an increasing number of particles occupying
the non-condensed mode.
A system is called fragmented if there is more than one large
eigenvalue~\cite{Mueller06a}.

\begin{figure}
  \includegraphics[width=\columnwidth]{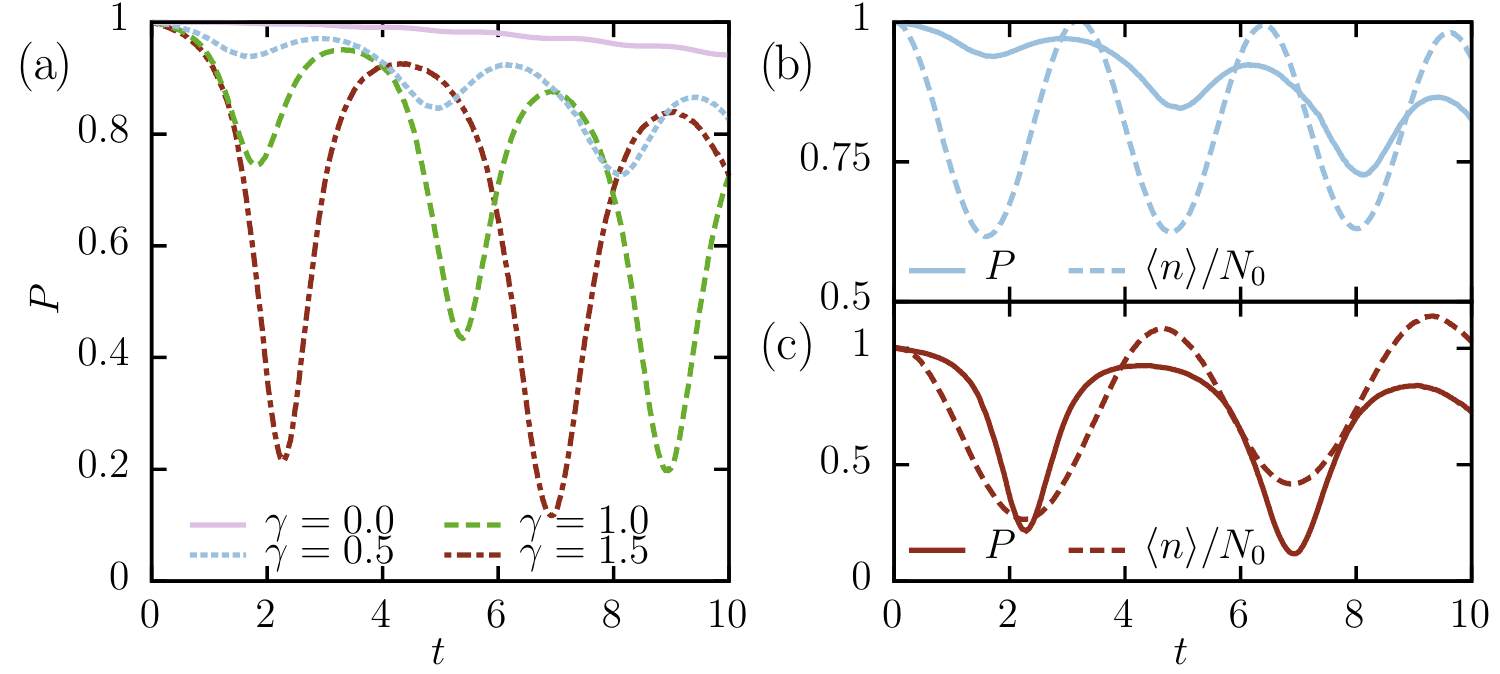}
  \caption{%
    (Color online)
    (a) Time evolution of the purity $P$ for four different values of the
    gain-loss parameter $\gamma$.
    With increasing values of $\gamma$ the oscillations become stronger and
    their frequencies become smaller.
    The comparison of the purity with the total particle number $\mean{n}$ for
    (b) $\gamma=0.5$ and (c) $\gamma=1.5$ shows the similar behavior of the two
    oscillation frequencies.
    In all calculations the pure initial state $c_{1/2} = 0.5 \pm 0.5i$ and the
    parameters $J = 1$, $g = 0.5$ and $N_0 = 100$ were used and it was
    averaged over 500 trajectories.
  }
  \label{fig:purityGamma}
\end{figure}

In Fig.~\ref{fig:purityGamma}(a) the time evolution of the purity is shown for
different values of the gain-loss parameter $\gamma$.
The remarkable feature here is that the purity shows oscillations.
The amplitude of these oscillations is heavily influenced by the strength of
the gain-loss parameter $\gamma$.
Tuning the gain-loss parameters to higher values results in much stronger
oscillations.
In the case $\gamma=1.5$, which is still significantly below the limit
$\gamma=2$ where the $\PT$-symmetric states vanish, the purity drops in its
first oscillation from $P=1$ to values as small as $0.2$ but then is nearly
fully restored to $P \gtrsim 0.9$.

In addition to the oscillations there is an overall decay of the purity.
Such a decay of purity also exists without gain and loss, $\gamma=0$.
In this case the purity will vanish but then is also restored due to the
elastic atomic interactions.
However, this revival takes place on much longer time scales (for the system
studied the first revival occurs at $t>100$).
Also this revival process is damped by particle losses~\cite{Pawlowski10a,
Sinatra98a} which stands in contrast to the purity oscillations discussed in
this work where the coupling to the environment is the driver behind the
oscillations.
Only if there is also no interaction between the particles, $g=0$, the
condensate will stay completely pure.

We notice that the oscillation frequency of the purity becomes smaller for
higher values of the gain-loss parameter $\gamma$.
It is known that in $\PT$-symmetric double-well systems the oscillations of the
total particle number show a similar behavior.
The total particle number oscillates as a result of the particle number
oscillations in each of the two wells.
For $\gamma=0$ the phase between the oscillations in the two wells is $\pi$ and
the total particle number is conserved, however, for an increasing gain-loss
parameter these oscillations become more and more in phase leading to the
oscillations of the total particle number~\cite{Klaiman08a}.
As can be seen in Figs.~\ref{fig:purityGamma}(b) and \ref{fig:purityGamma}(c)
the frequencies of the purity oscillations are in fact very similar to those of
the oscillations of the total particle number.
The minima and maxima of the two oscillations approximately coincide and, thus,
show a similar dependency on the gain-loss parameter.

\begin{figure}
  \includegraphics[width=\columnwidth]{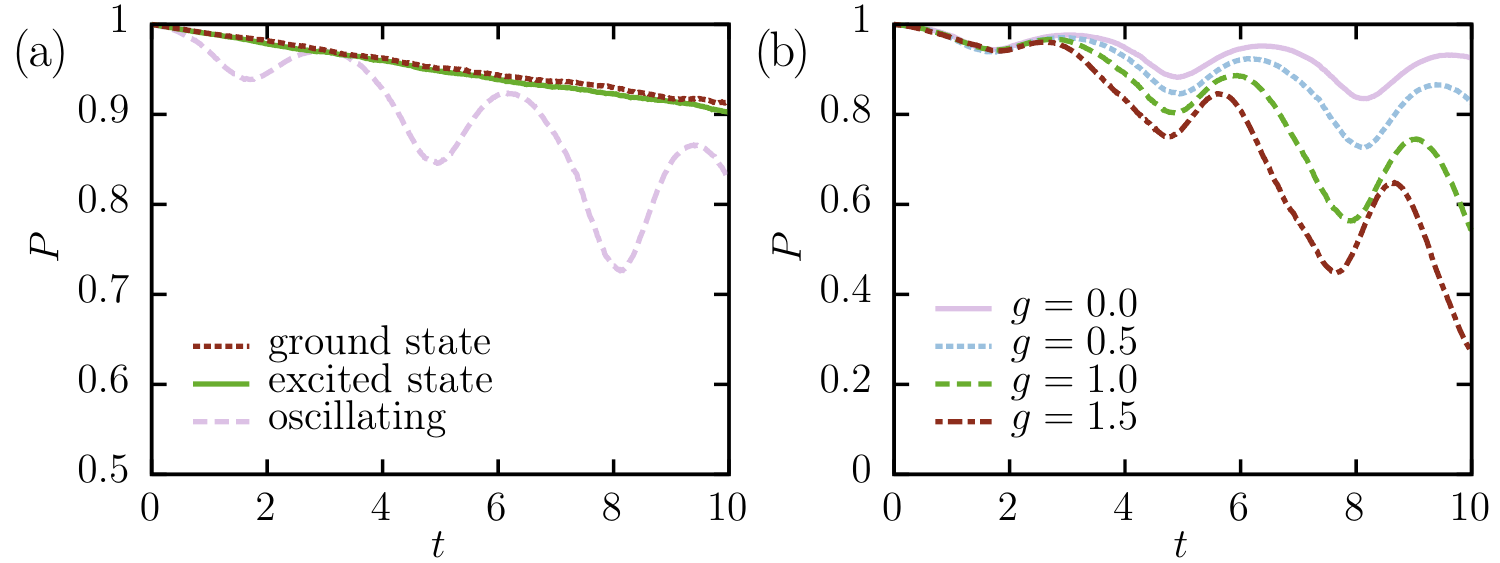}
  \caption{%
    (Color online)
    (a) The purity of the stationary ground and excited state of the
    $\PT$-symmetric Gross-Pitaevskii equation does not show oscillations but
    decays slowly, and in contrast to an oscillating state with $c_{1/2} = 0.5
    \pm 0.5i$ does not reach small purities.
    The results were obtained using $g = 0.5$.
    (b) Increasing the nonlinearity parameter $g$ results in a faster overall
    decay of the purity but changes only slightly the frequency of the
    oscillations.
    The initial state is pure with $c_{1/2} = 0.5 \pm 0.5i$.
    In all calculations the parameters $J = 1$, $\gamma = 0.5$ and $N_0 = 100$
    were used and it was averaged over 500 trajectories.
  }
  \label{fig:purityStatNonlin}
\end{figure}

The most prominent feature of $\PT$-symmetric systems is the existence of
stationary solutions despite the in- and outcoupling of particles.
Therefore, we are interested in the many-particle dynamics of the stationary
states of the $\PT$-symmetric Gross-Pitaevskii equation~\eqref{eq:discreteGPE}.
It has already been shown that the expectation values of the particle numbers
of the corresponding many-particle state also behave approximately stationary
when solving the time evolution with the master equation~\cite{Dast14a}.
Therefore, we do not expect oscillations of the purity in this case.
This is confirmed by Fig.~\ref{fig:purityStatNonlin}(a) where the purity of
the stationary ground and excited state is compared with an oscillating state
using the same gain-loss parameter $\gamma = 0.5$.
There are no oscillations of the purity for the stationary states but instead
the purity decays similar to the overall decay of an oscillating state.
As a result the stationary states stay almost pure, thus justifying the
use of the $\PT$-symmetric Gross-Pitaevskii equation to calculate stationary
solutions.

As a next step, we discuss the behavior of the purity for different values of
the nonlinearity parameter $g$ with the starting conditions of an oscillating
state.
Figure~\ref{fig:purityStatNonlin}(b) shows the time evolution of the purity for
different values of $g$ but with an identical gain-loss parameter $\gamma =
0.5$.
The increasing nonlinearity parameter $g$ results in a slightly higher
oscillation frequency.
This is not surprising since we have already seen that the oscillations of the
purity are similar to those of the total particle number and we know from
$\PT$-symmetric double-well systems that a greater nonlinearity slightly
increases the pulsing frequency~\cite{Dast13a}.
The main effect of the nonlinear term is that it increases the strength of
the overall decay of the purity.
This effect is not exclusive to systems with balanced gain and loss but also
occurs without coupling to the environment in the limit $\gamma=0$.

The purity oscillations have a direct impact on the spatial coherence between
the two lattice sites which can be measured by the average contrast
in interference experiments.
Such experiments can be realized by turning off the double-well trap whereby
the condensate expands and interferes~\cite{Gati06b, Shin04a}.
The average contrast is given by~\cite{Witthaut08a, Witthaut09a}
\begin{equation}
  \nu = \frac{2 |\mean{\ha_1^\dagger \ha_2}|}{\mean{\ha_1^\dagger \ha_1} +
  \mean{\ha_2^\dagger \ha_2}}  \ \in [0,1]
  \label{eq:contrast}
\end{equation}
and the squared contrast can be written as
\begin{equation}
  \nu^2 = P - I,
  \label{eq:contrastSquared}
\end{equation}
where we defined the imbalance of the particle number in the two lattice sites
\begin{equation}
  I = \left( \frac{\mean{\ha_1^\dagger\ha_1} - \mean{\ha_2^\dagger\ha_2}}
  {\mean{\ha_1^\dagger\ha_1} + \mean{\ha_2^\dagger\ha_2}}
  \right)^2 \ \in [0,1].
  \label{eq:imbalance}
\end{equation}
Note that the average contrast has to be understood as an ensemble average,
i.e.\ it is obtained by averaging over various experiments and, thus, is
reduced if the phase fluctuates for different measurements.
The contrast in a single measurement, however, is only reduced by an imbalance
of the particle number~\cite{Castin97a, Gati06b}.

\begin{figure}
  \includegraphics[width=\columnwidth]{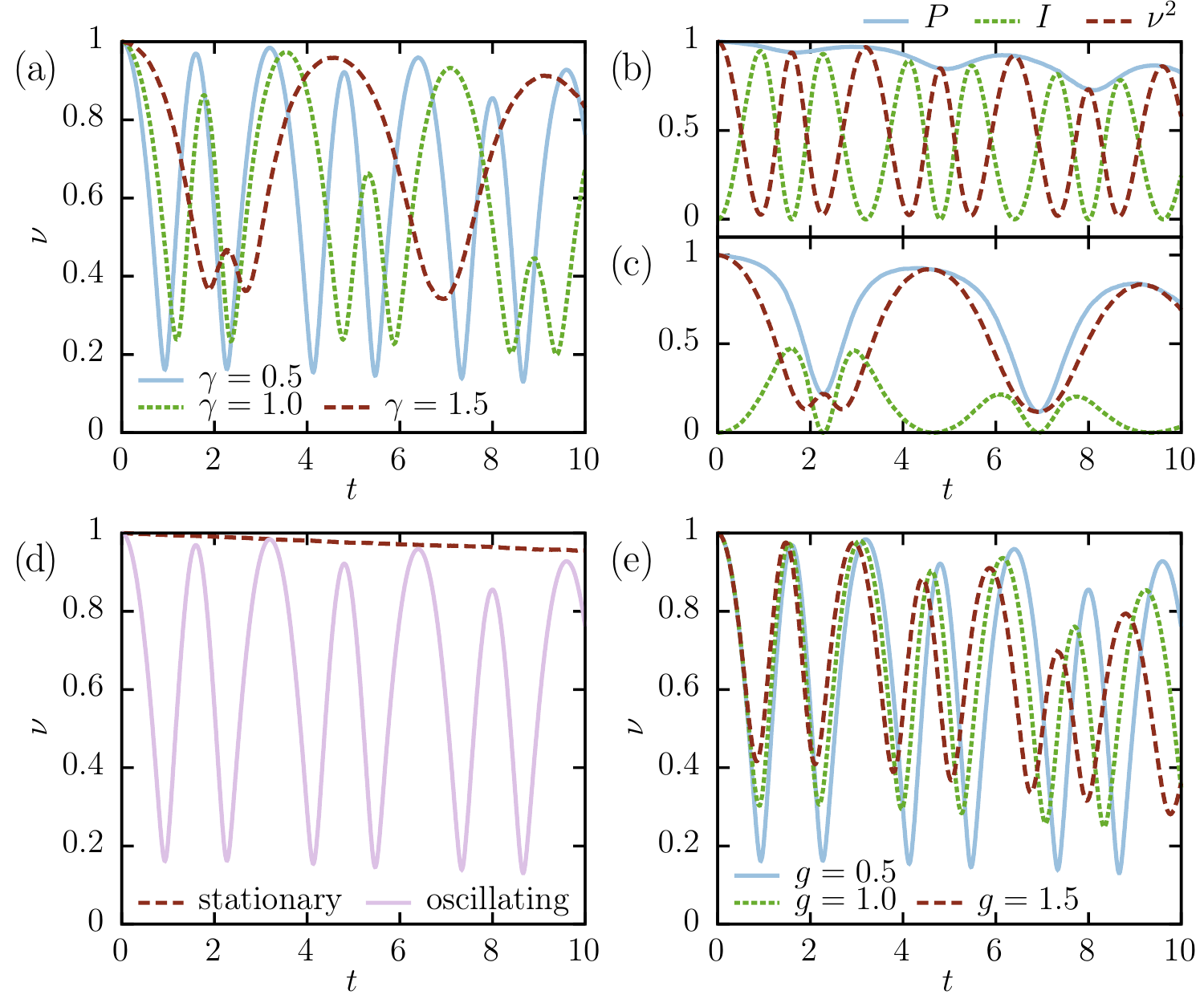}
  \caption{%
    (Color online)
    (a) The contrast $\nu$ for three different values of $\gamma$ shows
    oscillations.
    Every second peak is smaller and narrower.
    For (b) $\gamma = 0.5$ the overall behavior of the contrast is dominated by
    the imbalance $I$ and for (c) $\gamma = 1.5$ by the purity $P$.
    (d) The contrasts of the stationary ground and excited state of the
    $\PT$-symmetric Gross-Pitaevskii equation lie nearly perfectly on top of
    each other.
    They do not oscillate and stay high compared to an oscillating state.
    (e) The main effect of the nonlinearity $g$ is that it dampens the
    oscillations of the contrast.
    For all calculations except the stationary states the pure initial state
    $c_{1/2} = 0.5 \pm 0.5i$ was used.
    If not stated otherwise in the legend the parameters are $J = 1$, $\gamma =
    0.5$, $g = 0.5$ and $N_0 = 100$ and it was averaged over 500 trajectories.
  }
  \label{fig:contrast}
\end{figure}

The average contrast for different values of the gain-loss parameter $\gamma$
is shown in Fig.~\ref{fig:contrast}(a).
Since the initial wave function is pure, $P=1$, and the particles are evenly
distributed, $I=0$, the initial contrast is $\nu=1$.
Just as the purity oscillates, so does the contrast: It runs through small
values but then is nearly fully restored.
Every second peak is smaller and less broad, which can be seen very clearly for
$\gamma=1$.
For increasing values of $\gamma$ these peaks become smaller and can even
vanish.
This happens for $\gamma=1.5$, where the first small peak is still visible but
the second small peak has vanished.

This behavior can be understood by having a closer look at the components of
Eq.~\eqref{eq:contrastSquared}.
For the small value $\gamma = 0.5$ the purity, imbalance and squared contrast
is shown in Fig.~\ref{fig:contrast}(b).
The purity is the upper limit of the squared contrast.
The contrast is maximum where the imbalance reaches a minimum and vice versa.
Since the oscillations of the imbalance are stronger than the oscillations of
the purity at small values of $\gamma$ the overall behavior of the contrast is
dominated by the imbalance.

For the larger value $\gamma = 1.5$ the situation changes drastically as can be
seen in Fig.~\ref{fig:contrast}(c).
The purity oscillations are now strong enough to dominate the behavior of the
contrast.
Since the oscillations of the purity reach small enough values every second
peak of the contrast is either very small ($t\approx 2.25$) or even suppressed
($t \approx 7$).
The remaining peaks that coincide with the maxima of the purity become broader.

Note that in both cases the purity is at an extremum whenever the particles are
equally distributed, i.e.\ $I=0$.
This is a result of the already discussed observation that the extrema of
the purity coincide with the extrema of the total particle number.
Since for $I=0$ the contrast is not reduced by the imbalance, this allows a
precise measurement of the purity's extrema.

As discussed previously the purity of the stationary ground and excited state
of the $\PT$-symmetric Gross-Pitaevskii equation do not oscillate but only
decay slowly.
Also for $\PT$-symmetric states $c_1 = c_2^*$ holds, i.e.\ the particles are
equally distributed, $I=0$.
Thus we expect that the contrast stays high and does not oscillate which is
confirmed by Fig.~\ref{fig:contrast}(d).
For comparison an oscillating state is shown.

Looking at the contrast for different values of the nonlinearity in
Fig.~\ref{fig:contrast}(e) shows that the overall behavior of the contrast does
barely change.
The main effect of the interaction is that the amplitude of the oscillations
become smaller and the oscillation frequency is slightly increased.

To conclude we have shown that the purity undergoes oscillations, i.e.\
starting with an initially pure state the purity drops to small values but then
is almost completely restored.
This behavior is periodically repeated and the oscillations of the purity were
found to be in phase with the oscillations of the total particle number.
Tuning the strength of the gain and loss or the on-site interaction of the
atoms strongly influences both the amplitude and the frequency of the
oscillations.
Using the stationary states of the $\PT$-symmetric Gross-Pitaevskii equation as
initial states of the master equation does not yield such oscillations but the
condensate's purity decays only slowly.
The oscillations of the purity have a direct impact on the average contrast
that can be observed in interference experiments.
Since the purity is minimum or maximum at precisely the times where the
particles are equally distributed the purity's extrema can be directly measured
via the contrast.

Since the system studied in this work can be understood to be an elementary
building block of a larger transport chain~\cite{Kreibich13a, Kreibich14a} the
numerical study and the experimental accessibility~\cite{Gericke08a, Wurtz09a,
Robins08a, Doring09a, Schneble04a, Yoshikawa04a} of the information provide new
insight into transport processes of quantum systems ranging from Bose-Einstein
condensates to solid state systems, where the existence of macroscopic
superposition states can be decisive~\cite{Dounas-Frazer07a}.
Furthermore these results are important for the feasibility of continuous atom
lasers where the lasing mode is required to be pure~\cite{Robins08a} if the
atom laser is implemented with in- and outcoupling occurring at different
sites.
For future work it will be instructive to study the purity oscillations in an
analytical manner, e.g.\ by using the Bogoliubov backreaction
method~\cite{Vardi01a} to gain a deeper understanding of the underlying
process.

\end{document}